\documentstyle[12pt]{article}

\newcommand{\be}{\begin{equation}}
\newcommand{\ee}{\end{equation}}

\newtheorem{teo} {Theorem} [section]
\newtheorem{defi}[teo] {Definition} 
\def\bkC{{\rm \kern.24em 
\vrule width.05em 
\vrule height1.4ex 
depth-.05ex 
\kern-.26em C}}

\input{epsf}

\begin{document}

\setlength{\oddsidemargin}{0cm} \setlength{\baselineskip}{7mm}

\begin{normalsize}\begin{flushright}
\end{flushright}\end{normalsize}

\begin{center}
  
\vspace{25pt}
  
{\Large \bf Wormholes in Simplicial Minisuperspace }

\vspace{40pt}
  
{\sl Crist\'{o}v\~{a}o Correia da Silva}
$^{}$\footnote{e-mail address :
clbc2@damtp.cam.ac.uk}
and {\sl Ruth M. Williams}
$^{}$\footnote{e-mail address : rmw7@damtp.cam.ac.uk}
\\

\vspace{20pt}

DAMTP, Silver Street, Cambridge, \\
CB3 9EW, England \\

\end{center}

\vspace{30pt}

\begin{center} {\bf ABSTRACT } \end{center}
\vspace{12pt}
\noindent

We consider a simplicial minisuperspace modelling wormhole
configurations. Such configurations result from a cut and paste
procedure on two identical copies of the most relevant triangulation
of our universe, namely the cone over $\alpha_{4}$, which is the
simplest triangulation of $S^{3}$. The connecting space  where the two
copies are glued, acts as the throat of a wormhole connecting two
distinct universes. We show that the model  contains both Euclidean and
Lorentzian  classical configurations, although the Euclidean
configurations are restricted to very small size, while the Lorentzian
ones exist for all sizes. By computing the steepest descents contours
associated with the different kind of configurations, and studying the
behaviour of the Euclidean action along these contours we are able to
conclude that except for a very small range of boundary data, the
semiclassical approximation is always valid, which facilitates both
the computation and interpretation of the wavefunctions associated to
those steepest descents contours.
Having computed these semiclassical wavefunctions we could observe
that in the case of the microscopic Euclidean wormhole configurations there is a
strong prediction of a finite throat, which seems to indicate that
such configurations are very likely. For large universe there are only
Lorentzian configurations. Some of them have fully complex Euclidean action 
and the $exp(-ReI)$ acts as the weighting of the Lorentzian
configurations. We find large universes connected by similarly large
throats to be strongly favoured.

\vspace{24pt}

\vfill

\newpage

\section{Introduction}

In the last decade wormhole configurations have been extensively
studied and became more and more ``respectable'', as they were proved
to be almost ubiquitous. There are two main kind of wormholes. In 1988
Coleman , \cite{coleman},  proposed that the existence of microscopic Euclidean
wormholes could explain why the cosmological constant is so small in
our universe, and eventually yield a probability distribution for all
fundamental constants of nature. This launched a flurry of interest on
these kind of wormholes,(e.g. \cite{preskill}, \cite{hawkingwh}), although further results,(e.g. \cite{susskind}, \cite{gristein}), have cast
doubts on the initial  claims. On the other hand, macroscopic wormholes
connecting large Lorentzian universes have also been a subject of
intense study, e.g. \cite{hochberg}, \cite{hochbergvisser}.

In \cite{hochberg}, Hochberg considers a wormhole model that results from
taking two copies of (Lorentzian) Friedman-Robertson-Walker universes,

\be 
ds^{2}=-dt^{2}+R^{2}(t)\biggl\{\frac{dr^{2}}{1-kr^{2}}+r^{2}(d\theta
^{2}+\sin{\theta}^{2}\,d\phi ^{2})\biggr\}
\ee
and remove from each of them a $4-$dimensional region of the form
$\Omega_{1,2}=\{r_{1,2}\leq a\}$. The resulting $4-$spaces will have
identical boundaries $\partial \Omega_{1,2}=\{r_{1,2}=a\}$. By identifying these
two boundaries one obtains two FRW spacetimes connected by a wormhole,
whose throat is located on their common  $\partial \Omega$
boundaries.

 There is no need to invoke any
arbitrary $3+1$ decomposition of spacetime, usually ADM, as in the case of
continuum models. This means that global issues like topology can
still be addressed in the simplicial minisuperspace, which is not
possible in the continuum versions.

The  simplicial framework  deals with the
 universe in a fully $4-$D way,   not relying on any
 specific $3+1$ decom\-position of spacetime, usually ADM in the case
 of continuum models. This means that global issues like topology can
still be addressed in the simplicial minisuperspace, which is not
possible in the continuum versions. Additionally, it will allow us to use
 the same simplicial mini\-super\-space model to study both kinds of
 wormholes.   Simplicial minisuperspace models have proved very
 reliable in the past by  confirming results obtained with the
 continuum formalism \cite {h2}, \cite{birm}, \cite{cris2}. However, they are at their  best when applied to
 situations  which the continuum theory cannot easily handle. Namely,
 in the study of the effects of generalising the definition of history
 in quantum gravity, to include conifolds and not just
 manifolds. See \cite{hartle},\cite{witt2}, \cite{cris1}. Furthermore, they
 can sometimes deliver some unexpected results, like the existence of
 classical Euclidean solutions for all sizes of the boundary
 $3-$universe, when we consider an arbitrary scalar coupling
 $\eta R\phi^{2}$, \cite{cris3}.

\section{Simplicial Quantum Gravity}

The simplicial framework in QG is usually seen as an approximation to
 the more fundamental continuum framework. However, it has been
argued that at the scales where quantum gravity effects become
relevant, it is reasonable to assume that not just energy but even
space and time will be discrete. In this way the simplicial framework
could even be the  more fundamental one.

The basic building block of any simplicial complex is the $n-$simplex.

\begin{defi}
Given $n+1$ points in $R^{n+1}$, labelled $v_{1},v_{2},...,v_{n+1}$, if
these points are affinely independent, then an $n-$dimensional simplex
$\sigma^{n}$ is the convex hull of these points

\be
\sigma^{n}=\{x\in R^{n+1}:\,x=\sum_{i=1}^{n+1}\lambda_{i}v_{i} \,\,where
\,\,\lambda_{i}\geq 0 \,\,and\,\, \sum_{i=1}^{n+1}\lambda_{i}=1\}
\ee
\end{defi}

Thus, a $0-$simplex is a vertex, a $1-$simplex is an edge, a
$2-$simplex a triangle, etc. Any  subset  of these vertices also spans
a simplex, of lower dimension, which is called a face of $\sigma^{n}$.

From these building blocks we can construct a whole variety of spaces
in particular some which by obeying  certain natural conditions offer the
widest reasonable framework in which to investigate the concept of
simplicial geometry. These are called  simplicial complexes. 

\begin{defi}
A simplicial complex $(K,\mid K\mid )$ is a topological space $\mid K\mid$ and a
 collection of simplices $K$, such that

\begin{itemize}

   \item $\mid K\mid$ is a closed subset of some finite dimensional
   Euclidean space.

   \item  If $\sigma $ is a face of a simplex in $K$, then $\sigma $
   is also contained in $K$.

   \item  If $\sigma _{a}$ and $\sigma _{b}$ are simplices in $K$, then  
   $\sigma _{a} \cap \sigma _{b}$ is a face of both  $\sigma _{a}$
   and $\sigma _{b}$.

   \item The topological space $\mid K\mid $ is the union of all
   simplices in $K$.
\end{itemize}     
\end{defi}

By definition the dimension of the simplicial complex is the maximum
dimension of any of its simplices.

Each simplex in $K$ is totally
described by its vertices. 
Since a simplicial complex describes both
the simplices of the space $\mid K \mid$ and gives the rules for how
these building blocks are connected, then the simplicial complex
itself is uniquely determined by the vertices and the  rules stating
in which simplices they are contained in.

There are several standard topological definitions that extend
naturally to simplicial complexes, like

\begin{defi}
A simplicial complex is compact it it contains a finite number of
simplices.
\end{defi}
\begin{defi}
A simplicial complex is connected if any two of its vertices are
connected by a sequence of edges.
\end{defi}

 We shall see that in order for it to be possible to
define curvature in a complex, (and consequently an action functional
for QG), it is essential that this complex be pure, i.e.,

\begin{defi}
A simplicial $n-$complex, $(K^{n},\mid K^{n} \mid)$, is pure when every
lower dimensional simplex in it is contained in at least one
$n-$simplex of  $K^{n}$.
\end{defi}

 A natural definition of boundary is also essential. However,
it can be shown that the notion of boundary of a complex is meaningful for only a special subset
of pure simplicial complexes, which is defined by 

\begin{defi}
A pure simplicial $n-$complex $(K^{n},\mid K^{n} \mid)$, is said to
be non-branching if every $(n-1)-$simplex is contained in either one
or two $n-simplices$.
\end{defi}

Only now can we define the boundary of a complex as:

\begin{defi}
The boundary, $\partial K^{n-1}$, of a pure non-branching simplicial
 $n-$complex $K^{n}$ is the complex made up 
of all $(n-1)-$simplices that are faces of one and only one
$n-$simplex of $K^{n}$.
\end{defi}

An essential concept for the study of the relationship between
continuum spaces and their simplicial representations is that of
triangulation

\begin{defi}
A simplicial complex  $(K,\mid K\mid )$ is said to triangulate a
smooth space ${\cal{K}}$ if there is an homeomorphism $\Phi$ between
$\mid K\mid $ and   ${\cal{K}}$.
\end{defi}

Conventionally, if ${\cal{K}}$ is an $n-$sphere then it is
said that $(K,\mid K\mid )$ is a simplicial $n-$sphere, the same for an
$n-$ball, etc.  In the
topological sense we can say that $K$ is
the simplicial counterpart of ${\cal{K}}$.  

 What characterises a continuum
topological manifold is the fact that all its points have neighbourhoods
homeomorphic to open sets in $R^{n}$, like an open ball $B^{n}$.
 So the definition of the simplicial version of a continuum manifold
is dependent on the definition of the simplicial equivalent of
boundary of a point.  Only then can  the local topology of the
complexes be studied.

\begin{defi}
Given a simplicial complex $(K^{n},\mid K^{n} \mid)$, the
combinatorial star of a vertex $v\in K^{n}$, denoted $St(v)$, is the
complex consisting of all the $n-$simplices of $K^{n}$ that contain
$v$.
\end{defi}

\begin{defi}
Given a simplicial complex $(K^{n},\mid K^{n} \mid)$, the
combinatorial link of a vertex $v\in K^{n}$, denoted $L(v)$, is the
complex consisting of all the simplices that are in  $St(v)$ but do
not contain $v$ itself.
\end{defi}

The star of a vertex is very much like the neighbourhood of a point in
a continuum framework. So we expect that the defining characteristic of the
simplicial analogues of manifolds to  be that the star of each vertex be
homeomorphic to an open ball $B^{n}$, which is equivalent to say that
its star is a simplicial open $n-$ball. However it is
customary to reformulate these conditions in terms of links and not
stars. 

It is easy to see that to say that $St(v)$ is a simplicial open $n-$ball is
equivalent to say that $L(v)$ is a simplicial $(n-1)$-sphere.
And so we will restate the ``manifold condition'' as that the combinatorial link of every
vertex be homeomorphic to an $(n-1)$-sphere, i.e., $L(v)$ be a
simplicial $(n-1)-$sphere. 

For technical reasons the previous conditions are still not enough for
the resulting simplicial complexes to have the same properties as
continuum manifolds. One last condition is needed.

\begin{defi}
A simplicial $n-$complex $(K^{n},\mid K^{n} \mid)$ is said to be
strongly connected if any two simplices of $K^{n}$ can be connected by
a sequence of $n-$simplices each intersecting along some
$(n-1)$-simplex.
\end{defi}

We can now define the simplicial counterparts of smooth manifolds:

\begin{defi}
 
A combinatorial $n-$manifold ${\cal M}^{n}$, is an $n-$dimensional simplicial complex such  that  
       
\begin{itemize}

     \item It is pure.
     \item It is non-branching.
     \item Any two $n-$simplices can be connected by a sequence of $n-$simplices, each intersecting along some $(n-1)-$simplex.
     \item  The link of every vertex is a simplicial $(n-1)-$sphere.

\end{itemize}
\end{defi} 

Note that there are simplicial complexes that are homeomorphic to topological manifolds but are not combinatorial manifolds. The definition of combinatorial manifold carries more structure than simply the topology.

\subsection{Simplicial Geometry}

We have so far been concerned solely with the topological and combinatorial aspects of
 simplicial quantum gravity. We shall now focus on the geometric issues.

The simplest way to describe a geometry on an any pure non-branching simplicial complex
$K^{n}$ is to use the Regge formalism:

\begin{itemize}

\item 1)We require the metric on the
interior of each $n-$simplex of $K^{n}$ to be flat.

\item 2) We then assign
edge lengths to each edge of $K^{n}$. Not every assignment of edge
lengths is consistent with the simplices having flat interiors. The
triangle inequalities and their higher dimensional analogues must be
satisfied. Necessary and sufficient conditions for this to happen are that the
squared volumes of all $k-$simplices, with $k=2,3,...n$, must be
positive.

\end{itemize}
  
The metric information, which in the continuum spaces is contained
on the metric tensor $g_{\mu\nu}$, is 
imprinted on  a simplicial complex, $K^{n}$,  via an assignment of its squared edge lengths. 

\be
g_{\mu\mu}(x)\rightarrow
g_{ij}(\{s_{k}\})=\frac{s_{0i}+s_{0j}-s_{ij}}{2}
\ee 
where $0$ is just an arbitrary vertex of  $K^{n}$,
and $s_{ij}$ is the square edge length of the edge $[ij]$.

Note that in the simplicial framework the metric degrees of freedom
are now the squared edge lengths $s_{k}$. This means that all
geometric quantities should be expressible in terms of them.

Since by definition the metric inside each $n-$simplex is flat the
curvature in a simplicial complex cannot reside there. Instead it
is located at the $(n-2)-$simplices of the complex. The curvature
associated with each $(n-2)-$simplex is measured by what is called its
deficit angle, whose expression is slightly  different according to
whether the   $(n-2)-$simplex is located in the interior or on the
boundary of the complex:

For an interior  $(n-2)-$simplex, $\sigma_{i}^{n-2}$, the deficit
angle is 
\be
\theta(\sigma ^{n-2}_{i})=2\pi -\sum _{\sigma ^{n}\in St(\sigma ^{n-2}_{i})}\theta _{d}(\sigma ^{n-2}_{i},\sigma ^{n})   
\ee
where $\theta _{d}(\sigma ^{n-2}_{i},\sigma _{n})$ is called the
dihedral angle of $\sigma_{i}^{n-2}$ associated with
$\sigma_{i}^{n}$, and is the angle between the the two
$(n-1)$-simplices that belong to $\sigma_{i}^{n}$ and intersect at
$\sigma_{i}^{n-2})$.

For a  $(n-2)-$simplex,  $\sigma_{b}^{n-2}$, on the boundary of the
complex, the deficit angle is

\be
  \psi (\sigma ^{n-2}_{b})=\pi -\sum _{\sigma ^{n}\in St(\sigma ^{n-2}_{b})}\theta _{d}(\sigma ^{n-2}_{b},\sigma ^{n})    
\ee

It is easy to see why the previous definitions of curvature are only valid for pure
complexes.  Only in a pure complex can we be sure that any
$(n-2)-$simplex is indeed contained in some  $n-$simplex.

Following Regge, the continuum Einstein action can be discretized for
any pure non-branching complex as

\begin{eqnarray}
 I[K^{n},\{s_{k}\}]&=&\frac{-2}{16\pi G}\sum _{\sigma ^{n-2}_{i}} V_{n-2}(\sigma ^{n-2}_{i})\theta (\sigma ^{n-2}_{i}) +  {\frac{2\Lambda }{16\pi G}} \sum _{\sigma ^{n}}V_{n}(\sigma ^{n})\nonumber \\
                                      &-& \mbox{} {\frac{2}{16\pi G}}\sum _{\sigma ^{n-2}_{b}} V_{n-2}(\sigma ^{n-2}_{b})\psi (\sigma ^{n-2}_{b})
\end{eqnarray}
where $V_{p}(\sigma ^{p})$ is the $p-$ volume of the $p-$simplex,
 $\sigma _{p}$.

 It is easy to see that all the volumes as all the
 deficit angles can be written as functions of the squared edge
 lengths $\{s_{k}\}$. See \cite{birm} for their explicit form.

We can now write the wavefunction of the universe in simplicial
quantum cosmology as  

\be
\Psi [\partial {K}^{n-1},\{s_{b}\}]=\sum _{K^{n}}\int_{C} D\{s_{i}\}e^{-I[K^{n},\{s_{i}\},\{s_{b}\}]}  
\label{PsiSimpl}
\ee
where

\begin{itemize}

\item $ \{s_{i}\}$ are the squared lengths of the interior edges 

\item  $\{s_{b}\}$ are the squared lengths of the boundary  edges

\item The kind sum over complexes $K^{n}$ depends on the boundary
conditions one adheres to. For example the equivalent of
Hartle-Hawking`s no boundary proposal would be to consider a sum over all compact combinatorial $n-$manifolds
${\cal{M}}^{n}$ whose only boundary is $\partial {\cal{M}}^{n-1}$, and
over all simplicial $n-$geometries $ \{s_{k}\}$ that have boundary edges with
squared lengths  $\{s_{b}\}$ 

\end{itemize}

Note that the while in the continuum we had a functional metric
integral, we now have a well defined product of integrals over edge
lengths. So it seems we have removed the problems related to issues
of gauge fixing and renormalisation associated with defining the
measure of the space of metrics. However, such problems reappear in
any attempt to take the continuum limit of this Regge
integral. Similarly the functional integral associated with the  the
scalar field now becomes a product of well defined simple integrals
over the value of the field in each interior vertex.

\section{Simplicial Minisuperspace}

A simplicial minisuperspace approximation (in a pure gravity model) 
consists of imposing two kinds of restrictions on the quantities being
summed over  in $(\ref{PsiSimpl}$). Namely, it involves singling out a
particular complex or family of complexes and singling out a few edge
lengths, by making all others a function of them. We are thus imposing
restrictions of a topological and geometrical nature.

In our case we will have:
$$\sum _{{K}^{n}}\longrightarrow W^{4}$$
$$\{s_{i}\}\longrightarrow s_{i}$$
$$\{s_{b}\}\longrightarrow {a,b}$$

Where $W^{4}$ and the other restrictions will be explained in more
detail in the next section.

\subsection{Topology Restrictions}

Consider two identical copies of the combinatorial manifold
${\cal{C}}=0*\alpha_{4}$ we used in \cite{cris3} as a model of our
universe. We shall denote them ${\cal{C}}_{1}$, and
${\cal{C}}_{2}$. As we have shown in the previous chapter
these models provide a good approximation to the wavefunction of the
universe, by predicting classical Lorentzian spacetime in the late
universe. So we can consider them two separate but identical
$4-$D universes.

 We now proceed to construct a connection between them
acting as a wormhole connecting these two universes. To do so note
that ${\cal{C}}_{1}$ is composed of five
$4-$simplices:

$${\cal{C}}_{1}= [01234], [01345], [01235], [01245], [02345]$$

\vspace{18pt}

 Remember that the $4-$simplex $[12345]$ does not belong ${\cal{C}}_{1}$
because the triangulation $\alpha_{4}$ of the $3-$sphere, is just the
surface of $[12345]$.

 Similarly ${\cal{C}}_{2}$ is composed of five
equivalent $4-$simplices,

 $${\cal{C}}_{2}= [0'1'2'3'4'], [0'1'3'4'5'], [0'1'2'3'5'], [0'1'2'4'5'], [0'2'3'4'5']$$

These two copies of our model universe are displayed in figure $5.1$. 

If we now remove one of these identical
$4-$simplices from each complex, say, $[02345]$ and $[0'2'3'4'5']$,
and then
glue the resulting complexes through the identification of the vertices
$0=0'$, $2=2'$, $3=3'$, $4=4'$ and $5=5'$, then we have connected the
two independent  $4-$D universes, through a $3-$D throat, thus
modelling a wormhole. The resulting $4-$complex shall be denoted as
$W^{4}$, and the identified vertices will be denoted as $0,2,3,4,5$,
see figure $5.2$.
 There are of course two non-identified vertices $1$ and $1'$.
Note also that given the identification $0=0'$, the complex $W^{4}$
 has only one interior vertex. $W^{4}$ is composed of eight
$4-$simplices:

$$W^{4}= [01234], [01235], [01345], [01245], [01'234], [01'235], [01'245], [01'345]$$

Note that the $4-$simplex $[02345]$ does not exist in $W^{4}$.
Furthermore, $W^{4}$ has only one boundary, the $3-$complex made up of
the following $3-$simplices:

\begin{eqnarray}
\partial W^{4}&=&[1234],[1'234],[1235],[1'235],[1245],[1'245],\nonumber \\
             & &\mbox{} [1345],[1'345].
\end{eqnarray}

The $3-$simplex $[2345]$ does not exist in $W^{4}$, since it belongs
only to $[12345]$, $[1'2345]$ and $[02345]$ and by definition $W^{4}$ does not
contain these $4-$simplices, and since we want $W^{4}$ to be pure and
non branching  all $3-$simplices in $W^{4}$ have to belong either
to one or two $4-$simplices. However, the four triangles that form
the surface of $[2345]$, i.e., $[234]$, $[235]$, $[245]$ and $[345]$, do belong to
$W^{4}$.

We can thus conclude that the  boundary $\partial W^{4}$ is a
combinatorial $3-$manifold. Indeed if we compute the links of its
vertices all are homeomorphic to $S^{2}$.

\be
L_{\partial W}(1)=L(1')=[234]\cup[235]\cup[245]\cup[345]
\ee
So we see that the links of the vertices $1$ and $1'$ are simply the
surface of a tetrahedron $[2345]$.

\be
L_{\partial W}(2)=[134]\cup[1'34]\cup[135]\cup[1'35]\cup[145]\cup[1'45]
\ee
Thus, the link of vertex $2$ is just the surface of an
hexahedron. The same is true for vertices $3$, $4$ and $5$. This is
only valid because $[2345]$ does not belong to $W^{4}$. If that was not
the case the link of vertex $2$ would not be homeomorphic to $S^{2}$
and the boundary  $\partial W^{4}$ would not be a $3-$manifold.

The throat $T^{3}$ of the wormhole is the $3-$D surface of the $4-$simplex
$[02345]$ without $[2345]$ . See figure $3$. It is a combinatorial $3-$manifold cons\-titu\-ted by
the $3-$simplices: 

$$T^{3}= [0234], [0345], [0235], [0245]$$

\subsection{Geometry}

For simplicity  and since there is still only one
interior vertex, we shall assume that all interior edge lengths are 
equal, and their squared values are $s_{i}$.

$$s_{01}=s_{01'}=s_{02}=s_{03}=s_{04}=s_{05}=s_{i}$$

 However, it is essential
that not all boundary edge lengths are the same if we want to separate
the evolution of the whole boundary $3-$universe from that of the wormhole's
throat. So we shall assume that all boundary edges  belonging to the
throat $T^{3}$ will
have squared lengths  $b$,

$$s_{23}=s_{24}=s_{25}=s_{34}=s_{35}=s_{45}=b$$
while all the other boundary edges  have
square edge length $a$.

\section{Minisuperspace Wavefunction}

Having defined our simplicial minisuperspace model 
 our objective is now   to evaluate the
wavefunction associated with it. Although the
$4-$complex is now somewhat more complicated we simplify the model  by not
 considering any matter sector. The general
 formulae obviously still apply. So the minisuperspace wavefunction
 will be given by

\be
\Psi [a, b]=\int _{C}Ds_{i}e^{-I [s_{i}, a, b]}  
\ee

To implement this expression we start by computing the Euclideanized
Regge action for the model. Since we have no matter sector we have simply

\begin{eqnarray}
I[W^{4},a, b]&=&\nonumber\frac{-2}{16\pi G}\sum _{\sigma _{2}^{i}} V_{2}(\sigma _{2}^{i})\theta (\sigma _{2}^{i}) +  {\frac{2\Lambda }{16\pi G}} \sum _{\sigma _{4}}V_{4}(\sigma _{4}) \\
             &-&\mbox{} {\frac{2}{16\pi G}}\sum _{\sigma _{2}^{b}} V_{2}(\sigma_{2}^{b})\psi (\sigma _{2}^{b})
\end{eqnarray}
where:

\begin{itemize}

 \item $\sigma _{k}$ denotes a $k-$simplex belonging to the set $\Sigma _{k}$ of all  $k-$simplices in  $W^{4}$.

\item $\theta(\sigma _{2}^{i})$, is the deficit angle associated with
the interior $2-$simplex $\sigma _{2}^{i}=[ijk]$

\item   $\psi (\sigma _{2}^{b})$ is the deficit angle associated with the boundary $2-$simplex $\sigma _{2}^{b}$

\item $V_{k}(\sigma _{k})$ for $k=2,3,4$ is the $k-$volume associated
with the $k-$simplex, $\sigma _{k}$.

\end{itemize}

As before all these quantities can be expressed in terms of the
squared edge lengths, $s_{i}$, $a$ and $b$, according to the
expressions in appendix $1$.

\begin {itemize}

\item With our choice of edge lengths, all eight $4-$simplices that compose
$W^{4}$ are of the same type, and their $4-$volume can be computed as

\be
 V_{4}(\sigma _{4})=\frac{b^{2}}{48}\sqrt{3\frac{as_{i}}{b^{2}}-\frac{s_{i}}{b}-\frac{3}{4}\frac{a^{2}}{b^{2}}}
\ee

\item There is only one  type of boundary $3-$simplex. The  eight of them,  can be denoted as  $[aaabbb]$, i.e., they have three edges of squared length $a$ and three
edges of squared length $b$. They all have the same volume. As an
example take $[1234]$, its volume is

\be
V_{3}(\sigma _{3}^{b})=\frac{\sqrt{3}\,b^{3/2}}{12}\sqrt{\frac{a}{b}-\frac{1}{3}}\ee

\item As for interior $3-$simplices there are  two types. Type I can be
described as $[aabs_{i}s_{i}s_{i}]$, of which $[0123]$ is an
example. There are twelve of them and they all have the same volume

\be
V_{3}(\sigma
_{3}^{Ii})=\frac{b^{3/2}}{12}\sqrt{4\frac{as_{i}}{b^{2}}-\frac{s_{i}}{b}-\frac{a^{2}}{b^{2}}}
\ee

There are four more  interior $3-$simplices, of type II, which are
$[0234]$, $[0235]$, $[0345]$, and $[0245]$. They all have the same
volume

\be
V_{3}(\sigma_{3}^{IIi})=\frac{\sqrt{3}\,\,b^{3/2}}{12}\sqrt{\frac{s_{i}}{b}-\frac{1}{3}}
\ee

\item We have twelve boundary triangles of the kind $[aab]$, which we shall
call type I, e.g., $[123]$   and they all have the same area

\be
V_{2}(\sigma _{2}^{Ib})=\frac{b}{2}\sqrt{\frac{a}{b}-\frac{1}{4}} 
\ee

and four boundary triangles of the kind $[bbb]$, type II, e.g., $[234]$ whose area is

\be
V_{2}(\sigma _{2}^{IIb})=\frac{\sqrt{3}}{4}b 
\ee

\item There are also two kinds of interior triangles. Type I includes
the ones of the kind $[as_{i}s_{i}]$, e.g., $[012]$. They all the same
area

\be
V_{2}(\sigma _{2}^{Ii})=\frac{a}{2}\sqrt{\frac{s_{i}}{a}-\frac{1}{4}} 
\ee

There are six more of type II, $[bs_{i}s_{i}]$, like $[023]$, whose
area is

\be
V_{2}(\sigma _{2}^{IIi})=\frac{b}{2} \sqrt{\frac{s_{i}}{b}-\frac{1}{4}}
\ee

\end{itemize}

For simplicity from now on we shall be working with  rescaled metric variables:

\be
 \eta =\frac{s_{i}}{b}
\ee
\be
 \alpha=\frac{a}{b}
\ee
\be
 T=\frac{H^{2}b}{l^{2}}
\ee

where $H^{2}=l^{2}\Lambda /3$, and $l^{2}=16\pi G$ is the Planck length. We shall work in units where $c=\hbar =1$.

Since we have two types of interior and boundary triangles we can
expect at least four different deficit angles.
The deficit angle associated with all eight interior triangles of type I 
is the same:

\be
\theta (\sigma _{2}^{Ii})=2\pi -3\arccos{\Biggl\{\frac{1}{2}\frac{(4\alpha -2)\eta-\alpha^{2}}{(4\alpha-1)\eta-\alpha^{2}}\Biggr\}}  
\ee
Similarly, all six interior triangles of type II have the same
deficit angle

\be
\theta (\sigma _{2}^{IIi})=2\pi -4\arccos{\Biggl\{\frac{2\eta-\alpha}{2\sqrt{3\eta-1}\sqrt{(4\alpha-1)\eta-\alpha^{2}}}\Biggr\}}
\ee

As for the deficit angles associated with the boundary triangles, we
can show that all twelve boundary triangles of type I have the same
deficit angle

\be
\psi (\sigma _{2}^{Ib})=\pi -2\arccos{\Biggl\{ \frac{\alpha}{2\sqrt{3\alpha-1}\sqrt{(4\alpha-1)\eta-\alpha^{2}}}\Biggr\}}
\ee
the remaining four boundary triangles of type II also  have the
same deficit angle

\be
\psi (\sigma _{2}^{IIb})=\pi -2\arccos{\Biggl\{\frac{(3\alpha -2)}{2\sqrt{3\alpha-1}\sqrt{3\eta-1}}\Biggr\}}
\ee

We can now write the explicit expression of the Regge action for this
model:

\begin{eqnarray}
I[\eta, T,\alpha]&=&-\frac{T}{H^{2}}\biggl\{8\sqrt{\alpha}\sqrt{\eta-\frac{\alpha}{4}}\biggl[2\pi -3\arccos{\biggl(\frac{1}{2}\frac{(4\alpha -2)\eta-\alpha^{2}}{(4\alpha-1)\eta-\alpha^{2}}\biggr)}\biggr] \nonumber\\
                  &+& 6\sqrt{\eta-1/4}\biggl[2\pi
                  -4\arccos{\biggl(\frac{2\eta-\alpha}{2\sqrt{3\eta-1}\sqrt{(4\alpha-1)\eta-\alpha^{2}}}\biggr)}\biggr] \nonumber \\
                  &+& 6\sqrt{4\alpha-1}\biggl[\pi
                  -2\arccos{\biggl(\frac{\alpha}{2\sqrt{3\alpha-1}\sqrt{(4\alpha-1)\eta-\alpha^{2}}}\biggr)}\biggr] \nonumber \\ 
                  &+&
                  2\sqrt{3}\biggl[\pi-2\arccos{\biggl(\frac{(3\alpha-2)}{2\sqrt{3\alpha-1}\sqrt{3\eta-1}}\biggr)}\biggr]   \\
                  &+&\mbox{} \nonumber \frac{T^{2}}{H^{2}}\biggl\{\alpha\sqrt{\frac{(3\alpha-1)\eta}{\alpha^{2}}-\frac{3}{4}}\,\,\biggr\}
\label{action1}
\end{eqnarray}

\subsection{Analytic Study of The Action}

For obvious physical reasons we require the boundary edge lengths $a$
and $b$, and thus $T$ and $\alpha$,  to be real and positive. 
On the other hand since we are only interested in geometries in which the
boundary three-metric is positive definite, we must require that the
volume of the boundary three-simplices be positive which is
equivalent to requiring that $\alpha > \frac{1}{3}$.

However, as  pointed out in  $\cite{hh}$, for it to be possible
for the wavefunction of
the universe to predict a classical Lorentzian Universe the
integration contour considered must be over complex metrics. Thus we
are lead to consider $\eta$ as being a complex variable, and the
analytic study of the action as a function  of a complex variable $\eta$
is essential.

The first thing to do is to identify the branch points of the
action. Terms like $\sqrt{z-z_{0}}$, are double-valued and have a
square-root branch point at $z=z_{0}$. In order to make this term
continuous we need to cut the complex plane. The most common branch
cut is simply $ (-\infty ,z_{0}]$.

So we see that the Euclidean
action has $3$ square-root branch points located at:
$$\eta_{0}=\frac{3}{4}\frac{\alpha^{2}}{3\alpha -1}$$
$$\eta_{2}=\frac{\alpha}{4}$$
$$\eta_{3}=\frac{1}{4}$$

On the other hand a term like $\arccos{u(z)}$, is infinitely
many-valued and has branch points at $u(z)=+1, -1$,
and at $u(z)=\infty$. The associated branch cuts are usually taken
to be $(-\infty ,-1]\cup [1,+\infty )$.

Since

$$\arccos{u(z)}=-i\log \biggl(u(z)+\sqrt{u(z)^{2}-1}\biggr)$$
then we see that there are logarithmic singularities when
$u(z)=\infty$. The table below  shows the logarithmic branching points and
infinities associated with the dihedral angles, where 

$$\eta_{1}=\frac{\alpha^{2}}{4\alpha-1}$$

\vspace{55pt}

\begin{tabular}{|c|c|c|c|}  \hline 
\hspace{.1in}Dihedral angles\hspace{.4in}&\hspace{.2in}$+1$\hspace{.2in}
  &\hspace{.2in}$-1$\hspace{.2in}
  &\hspace{.2in}$\infty $\hspace{.2in} \\ \hline
\hspace{.2in}$\theta (\sigma _{2}^{Ii})$\hspace{.2in}   &
\hspace{.2in}$1/4$, $\eta_{0}$ \hspace{.2in}      & \hspace{.2in}
$1/4$, $\eta_{0}$\hspace{.2in}  &\hspace{.2in}$1/3$\hspace{.2in}      \\ \hline

\hspace{.2in}$\theta (\sigma _{2}^{IIi})$\hspace{.2in}   & \hspace{.2in}$\alpha/4$\hspace{.2in}      & \hspace{.2in} $\eta_{0}$\hspace{.2in} &\hspace{.2in}$\eta_{1}$\hspace{.2in}        \\ \hline

\hspace{.2in}$\theta (\sigma _{2}^{Ib})$\hspace{.2in} & \hspace{.2in}$\eta_{0}$\hspace{.2in}      & \hspace{.2in} $\eta_{0}$\hspace{.2in} &\hspace{.2in}$\eta_{1}$\hspace{.2in}        \\ \hline
\hspace{.2in}$\theta (\sigma _{2}^{IIb})$\hspace{.2in} &
\hspace{.2in}$\eta_{0}$\hspace{.2in}      &
\hspace{.2in} $\eta_{0}$\hspace{.2in} &\hspace{.2in}$1/3$\hspace{.2in}        \\ \hline

\end{tabular}

\vspace{45pt}

In order to determine which branch cuts to take we need to know the
relative values of these critical points, but  those are dependent
on the value of the boundary parameter $\alpha$. There are however
some relations that are always valid for all $\alpha >1/3$

$$\eta_{0}> \frac{\alpha}{4},\frac{1}{3}$$
$$\eta_{1}> \frac{1}{4}$$
$$\eta_{0}>\eta_{1}$$
$$\eta_{1}> \frac{\alpha}{4}$$
So we see that the only indetermi\-na\-tes are
$min(\eta_{1},\frac{1}{3})$,  $min(\frac{\alpha}{4},\frac{1}{4})$
and  $min(\frac{\alpha}{4},\frac{1}{3})$. 

However, it is easy to see that the first two
are just the same $min(\alpha,1)$, and the third can of course be
rewritten as  $min(\alpha,\frac{4}{3})$. 

Thus, we have three different regions
\begin{itemize}

\item $\frac{1}{3}<\alpha< 1$, where
$\frac{\alpha}{4}<\frac{1}{4}<\eta_{1}<\frac{1}{3}<\eta_{0}$

\item $1<\alpha< \frac{4}{3}$, where
$\frac{1}{4}<\frac{\alpha}{4}<\frac{1}{3}<\eta_{1}<\eta_{0}$

\item $\frac{4}{3}<\alpha< +\infty $, where
$\frac{1}{4}<\frac{1}{3}<\frac{\alpha}{4}<\eta_{1}<\eta_{0}$

\end{itemize} 

So when we consider all the cuts associated to all the terms we shall
take as total branch cut $(-\infty, \eta_{0}]$. This also takes care
of the logarithmic singularity at $\eta=1/3$. 

This cut leads  to a continuous  action function, in
the complex plane with a cut $(-\infty ,\eta_{0}]$. However this function is still
infinitely many-valued. As in the previous chapters in order to remove
this multivaluedness we redefine the domain where the action is
defined, from the complex plane to the Riemann surface associated with
$I$. The infinite multivaluedness of the action reflects itself in $I$
having an infinite number of branches with taking different values. The
Riemann surface, ${\bf{R}}$ is composed by an infinite number of identical sheets, $\bkC -(-\infty ,\frac{3}{8}]$, one sheet for each branch of $I$.  

 The first sheet ${\bkC} _{1}$ of the action is defined  as the sheet
where the terms in $\arccos (z)$ assume their principal  values. That
is the sheet where the action takes the form ($\ref{action1}$), and so
$I^{I}[\eta,\alpha, T]=I[\eta,\alpha, T]$. 
When we continue the action in $\xi $ around one or more branch
points, we will leave this first sheet and emerge in some other sheet
of the Riemann surface. Only a few of these other sheets are relevant
to us. 

When we encircle all five branch points we leave the first sheet and
enter what we shall call the second sheet. It is easy to see that if we
cross the branch cut  $(-\infty ,\eta_{0}]$, somewhere between
$-\infty$ and the smallest branch point ($\alpha/4$ or $1/4$, according
to the value of $\alpha$) all terms of the action
change their sign in this new sheet. Thus, the action in the second sheet differs by an
overall minus sign relative to the action in the first sheet,
$I^{II}[\eta,\alpha, T]=-I^{I}[\eta,\alpha, T]$.

However, we shall see that the steepest descents (SD) contours yielding the
desired wavefunctions of the universe  cross the branch cuts
at other locations, thus emerging onto other sheets. For example, when
$\alpha >4/3$, if we take a contour that encircles the branch point
$1/4$, then at the first crossing of the branch cut, somewhere in
$(-\infty,1/4]$, we go from the first sheet ${\bkC} _{1}$
to the second ${\bkC} _{2}$, and  the action changes overall sign.
 Continuing in such a way  that we again cross the  branch cut, now between $1/4$ and the
nearest branch point $1/3$, the contour  enters what we shall call
the third sheet. By carefully studying the behaviour of each term at
the branch crossing, taking into account the branch cuts associated to
each of them, we conclude that in this third sheet, 
${\bkC} _{3}$, the action is

\begin{eqnarray}
I^{III}[\eta, T,\alpha]&=&-\frac{T}{H^{2}}\biggl\{8\sqrt{\alpha}\sqrt{\eta-\frac{\alpha}{4}}\biggl[2\pi -3\arccos{\biggl(\frac{1}{2}\frac{(4\alpha -2)\eta-\alpha^{2}}{(4\alpha-1)\eta-\alpha^{2}}\biggr)}\biggr]\nonumber \\
                  &+&- 6\sqrt{\eta-1/4}\biggl[2\pi
                  +4\arccos{\biggl(\frac{2\eta-\alpha}{2\sqrt{3\eta-1}\sqrt{(4\alpha-1)\eta-\alpha^{2}}}\biggr)}\biggr]\nonumber  \\
                  &+& 6\sqrt{4\alpha-1}\biggl[\pi
                  -2\arccos{\biggl(\frac{\alpha}{2\sqrt{3\alpha-1}\sqrt{(4\alpha-1)\eta-\alpha^{2}}}\biggr)}\biggr]\nonumber  \\ 
                  &+&
                  2\sqrt{3}\biggl[\pi-2\arccos{\biggl(\frac{(3\alpha-2)}{2\sqrt{3\alpha-1}\sqrt{3\eta-1}}\biggr)}\biggr]   \\
                  &+&\mbox{}\nonumber \frac{T^{2}}{H^{2}}\biggl\{\alpha\sqrt{\frac{(3\alpha-1)\eta}{\alpha^{2}}-\frac{3}{4}}\,\,\biggr\}
\end{eqnarray}

There is another important case, (still when $\alpha>4/3$), in which the SD contour encircles
another branch point, namely $\alpha/4$. In this case we are
confronted with a contour that starting on the first sheet, crosses the
branch cut between $1/3$ and $\alpha/4$, encircles  $\alpha/4$ and
crosses the cut, once again somewhere between  $\alpha/4$ and
$\eta_{1}$. During the first crossing if we pay attention to the
branch cuts in each term of the action, all terms of the action
change their signs except the one associated with
$\arccos{\theta(\sigma_{b}^{II})}$, and so we end up at what we shall
call the fourth sheet ${\bkC} _{4}$, where the action is 

\begin{eqnarray}
I^{IV}[\eta, T, \alpha]&=&\frac{T}{H^{2}}\biggl\{8\sqrt{\alpha}\sqrt{\eta-\frac{\alpha}{4}}\biggl[2\pi -3\arccos{\biggl(\frac{1}{2}\frac{(4\alpha -2)\eta-\alpha^{2}}{(4\alpha-1)\eta-\alpha^{2}}\biggr)}\biggr]\nonumber\\
                  &+& 6\sqrt{\eta-1/4}\biggl[2\pi
                  -4\arccos{\biggl(\frac{2\eta-\alpha}{2\sqrt{3\eta-1}\sqrt{(4\alpha-1)\eta-\alpha^{2}}}\biggr)}\biggr] \nonumber \\
                  &+& 6\sqrt{4\alpha-1}\biggl[\pi
                  -2\arccos{\biggl(\frac{\alpha}{2\sqrt{3\alpha-1}\sqrt{(4\alpha-1)\eta-\alpha^{2}}}\biggr)}\biggr]\nonumber  \\ 
                  &+&
                  2\sqrt{3}\biggl[-\pi-2\arccos{\biggl(\frac{(3\alpha-2)}{2\sqrt{3\alpha-1}\sqrt{3\eta-1}}\biggr)}\biggr]   \\
                  &-&\mbox{}\nonumber \frac{T^{2}}{H^{2}}\biggl\{\alpha\sqrt{\frac{(3\alpha-1)\eta}{\alpha^{2}}-\frac{3}{4}}\,\,\biggr\}
\end{eqnarray}

We then proceed to the second branch crossing somewhere between  $\alpha/4$ and
$\eta_{1}$, from  where we emerge onto yet  another sheet, ${\bkC} _{5}$. Once more if we pay attention to the
branch cuts in each term of the action we can conclude that the action
in this fifth sheet takes the form

\begin{eqnarray}
I^{V}[\eta, T, \alpha]&=&-\frac{T}{H^{2}}\biggl\{-8\sqrt{\alpha}\sqrt{\eta-\frac{\alpha}{4}}\biggl[2\pi +3\arccos{\biggl(\frac{1}{2}\frac{(4\alpha -2)\eta-\alpha^{2}}{(4\alpha-1)\eta-\alpha^{2}}\biggr)}\biggr]\nonumber\\
                  &+& 6\sqrt{\eta-1/4}\biggl[2\pi
                  -4\arccos{\biggl(\frac{2\eta-\alpha}{2\sqrt{3\eta-1}\sqrt{(4\alpha-1)\eta-\alpha^{2}}}\biggr)}\biggr] \nonumber \\
                  &+& 6\sqrt{4\alpha-1}\biggl[\pi
                  -2\arccos{\biggl(\frac{\alpha}{2\sqrt{3\alpha-1}\sqrt{(4\alpha-1)\eta-\alpha^{2}}}\biggr)}\biggr]\nonumber  \\ 
                  &+&
                  2\sqrt{3}\biggl[\pi-2\arccos{\biggl(\frac{(3\alpha-2)}{2\sqrt{3\alpha-1}\sqrt{3\eta-1}}\biggr)}\biggr]   \\
                  &+&\mbox{}\nonumber \frac{T^{2}}{H^{2}}\biggl\{\alpha\sqrt{\frac{(3\alpha-1)\eta}{\alpha^{2}}-\frac{3}{4}}\,\,\biggr\}
\end{eqnarray}

We have just examined two cases that will be relevant when dealing
with the SD contours, however, the lesson to be taken is that given
the high number of branch points, the Riemann surface of the action is
highly non-trivial and we must consider the changing form of the
action along its sheets.

\subsection{Asymptotic Behaviour of the Action}

Once more the asymptotic behaviour of the action with respect to the
integration variable, $\eta $, is essential to the study of the
convergence of the path integral yielding the wavefunction. Only after we
know this asymptotic behaviour can we guarantee that the steepest
descents contour is indeed dominated by the correct classical
solutions. However from what we have seen in the
previous chapter we must study this asymptotic behaviour not just in
the first sheet but in all others where the SD contours are liable to
go to infinity.

It is easy to see that in the first sheet, the asymptotic behaviour 
of the action is

\be
I^{I}[\eta\rightarrow \infty,\alpha,T]\sim \frac{\sqrt{3\alpha-1}}{H^{2}}T\biggl[T-T_{crit}^{I}(\alpha)\biggr]\sqrt{\eta}
\ee
where
\begin{eqnarray}
T_{crit}^{I}(\alpha)&=&\frac{8\sqrt{\alpha}}{\sqrt{3\alpha-1}}\biggl[2\pi-3\arccos{\biggl(\frac{2\alpha-1}{4\alpha-1}\biggr)}\biggr]\\
                    &+&\nonumber\mbox{} \frac{6}{\sqrt{3\alpha-1}}\biggl[2\pi-4\arccos{\biggl(\frac{1}{\sqrt{3}\sqrt{4\alpha-1}}\biggr)}\biggr];
\end{eqnarray}

for the third sheet we have

\be
I^{III}[\eta\rightarrow \infty,\alpha,T]\sim \frac{\sqrt{3\alpha-1}}{H^{2}}T(T-T_{crit}^{III})\sqrt{\eta}
\ee
where now
\begin{eqnarray}
T_{crit}^{III}(\alpha)&=&\frac{8\sqrt{\alpha}}{\sqrt{3\alpha-1}}\biggl[2\pi-3\arccos{\biggl(\frac{2\alpha-1}{4\alpha-1}\biggr)}\biggr]\\
                      &-&\mbox{}\nonumber\frac{6}{\sqrt{3\alpha-1}}\biggl[2\pi+4\arccos{\biggl(\frac{1}{\sqrt{3}\sqrt{4\alpha-1}}\biggr)}\biggr]
\end{eqnarray}
and finally for the fifth sheet
\be
I^{V}[\eta\rightarrow \infty,\alpha,T]\sim \frac{\sqrt{3\alpha-1}}{H^{2}}T(T-T_{crit}^{V})\sqrt{\eta}
\ee
where now
\begin{eqnarray}
T_{crit}^{V}(\alpha)&=&-\frac{8\sqrt{\alpha}}{\sqrt{3\alpha-1}}\biggl[2\pi+3\arccos{\biggl(\frac{2\alpha-1}{4\alpha-1}\biggr)}\biggr]\\
                    &+&\mbox{}\nonumber\frac{6}{\sqrt{3\alpha-1}}\biggl[2\pi-4\arccos{\biggl(\frac{1}{\sqrt{3}\sqrt{4\alpha-1}}\biggr)}\biggr].
\end{eqnarray}

In figure $4$ we see the plot of $T_{crit}^{I}(\alpha)$. It becomes infinite 
as $\alpha$ approaches $1/3$, which corresponds to the vanishing of
the volume of $\sigma _{3}^{Ib}$. However, for all other values of
$\alpha$ away from $1/3$ it quickly settles down at its asymptotic
value  $Tcrit^{I}(+\infty)=14.51$.

In the case of $T_{crit}^{III}$ it becomes $-\infty$ when
$\alpha\rightarrow 1/3$, and as $\alpha$ increases it becomes
positive, having the same asymptotic value as $T_{crit}^{I}$, which
is its upper bound. As for $T_{crit}^{V}$ it is always negative,
whatever the value of $\alpha$.

\section{Classical Solutions}

Since the complex $W^{4}$ is composed by eight equivalent
$4-$simplices, the metric of the complex will coincide with the metric
of each simplex.
The simplicial metric is then
\be
g_{ij}(\{s_{k}\})=\frac{s_{0i}+s_{0j}-s_{ij}}{2}
\ee 
where $i\neq j=1,2,3,4,5$.

If we now calculate the eigenvalues of $g_{ij}$, we get
$(\frac{1}{2},\frac{1}{2}, \lambda_{-},\lambda_{+})$, where

\be
\lambda_{+}=2\eta-\frac{1}{2}+\frac{1}{2}\sqrt{16\eta^{2}-4\eta+1-12\eta\alpha+3\alpha^{2}}
\label{lambda+}
\ee
\be
\lambda_{-}=2\eta-\frac{1}{2}-\frac{1}{2}\sqrt{16\eta^{2}-4\eta+1-12\eta\alpha+3\alpha^{2}}
\label{lambda-}
\ee

It is easy to see that since $\alpha>1/3$ then
$\lambda_{+}(\alpha,\eta)$ is always positive, whatever the value of $\eta$.

For $\lambda_{-}$, the situation is different.

\begin{itemize}
 
\item If $\eta>\eta_{0}$ then $\lambda_{-}(\alpha,\eta)>0$

\item If $\eta<\eta_{0}$ then $\lambda_{-}(\alpha,\eta)<0$

\end{itemize}

So we see that the  complex $W^{4}$ will have Euclidean signature
$(++++)$, if $\eta>\eta_{0}$, and will have Lorentzian signature
$(-+++)$ if $\eta<\eta_{0}$.

Since there is only one internal degree of freedom, $\eta$, there is
only one classical equation

\be
\frac{\partial I}{\partial \eta}=0
\ee
which takes the form

\begin{eqnarray}
T&=&8\sqrt{\frac{\alpha}{3\alpha-1}}\sqrt{\frac{\eta-\eta_{0}}{\eta-\alpha/4}}\biggl[2\pi-3\arccos{\biggl(\frac{1}{2}\frac{(4\alpha-2)\eta-\alpha^{2}}{(4\alpha-1)\eta-\alpha^{2}}\biggr)}\biggr]\\
 &+&\mbox{} \nonumber \frac{6}{\sqrt{3\alpha-1}}\sqrt{\frac{\eta-\eta_{0}}{\eta-1/4}}\biggl[2\pi-4\arccos{\biggl(\frac{2\eta-\alpha}{{2\sqrt{3\eta-1}}\sqrt{(4\alpha-1)\eta-\alpha^{2}}}\biggr)}\biggr]
\end{eqnarray}

Of course the classical solutions are of the form $\eta=\eta_{cl}(\alpha,T)$
where $T$ and $\alpha$ are boundary data. However, it is obvious
that   the previous expression does not lead to a closed expression of
$\eta$ as a function of $\alpha$ and $T$. So as in the previous
chapters we plot the classical solutions as if $T$ was the dependent
variable and not $\eta$. Of course this changes nothing, and it serves
only as a visual aid to understanding the solutions.

Note that  the asymptotic behaviour of $T(\alpha_{cl},\eta_{cl})$, when
$\eta_{cl}\rightarrow +\infty$ and  when $\eta_{cl}\rightarrow -\infty$, is the same

\begin{eqnarray}
T(\eta_{cl}\rightarrow \infty)=T_{crit}^{I}(\alpha)&=&\frac{8\sqrt{\alpha}}{\sqrt{3\alpha-1}}\biggl[2\pi-3\arccos{\biggl(\frac{2\alpha-1}{4\alpha-1}\biggr)}\biggr]\\
                    &+&\mbox{} \nonumber\frac{6}{\sqrt{3\alpha-1}}\biggl[2\pi-4\arccos{\biggl(\frac{1}{\sqrt{3}\sqrt{4\alpha-1}}\biggr)}\biggr]
\end{eqnarray}

As in the previous chapters, and for exactly the same reasons, all
classical solutions occur in pairs. Each element of the pair
$(\eta_{cl}^{I}, \eta_{cl}^{II})$, is numerically equal
$\eta_{cl}^{I}= \eta_{cl}^{II}$, but is located respectively in the
first and second sheets of the action, and so they have Euclidean
actions of opposite sign
$I^{I}[\eta_{cl}^{I}]=-I^{II}[\eta_{cl}^{II}]$.

For physical reasons we only consider solutions for which $T$ is
real positive and $\alpha\in [1/3, +\infty)$.
By plotting the solutions we conclude that there are three different
cases according to the value of $\alpha$.

\begin{itemize}

\item CASE 1 corresponds to $\frac{1}{3}<\alpha<1$.

As we can see in figure $5$ , in this interval there are
real Lorentzian solutions for $\eta\in (-\infty ,\alpha/4]$, and real Euclidean
solutions for $\eta\in [\eta_{0}, +\infty]$. Note however that
although it is not apparent in  figure $5$ the Euclidean solutions only
obey the physical condition $T>0$ from a value of $\eta$ slightly
larger than $\eta_{0}$. For
$\eta\in[\alpha/4,\eta_{0}]$ and for all non-real values of $\eta$ the
solutions have complex geometry.
 The divergence of
$T(\eta_{cl},\alpha)$ at $\alpha/4$ guarantees that there are Lorentzian
solutions for all values of $T>T_{crit}^{I}$.

On the other hand, we have classical Euclidean
solutions for all values between $0$ and  $T_{crit}(\alpha)$. So the existence of Euclidean solutions is limited by a ceiling
$T=T_{crit}^{I}(\alpha)$.
  However, since $T_{crit}^{I}(\alpha\rightarrow
1/3)\rightarrow +\infty$,  in this range of $\alpha$ we can pretty much
envisage raising the ceiling as much as we want, and so there will be
Euclidean classical solutions for every value of $T$, it is just a
case of getting $\alpha$ ever closer to $1/3$. Furthermore, note that
as $\alpha \rightarrow  1/3$  the volume of the boundary tetrahedra 

$$V_{3}[\sigma_{3}^{b}]=\frac{\sqrt{3}l^{3}}{12H^{3}}\,T_{crit}^{3/2}(\alpha)\sqrt{\alpha-1/3}$$
does not vanish because $T_{crit}(\alpha\rightarrow 1/3)\sim (\alpha-1/3)^{-1/4}$.

Regarding the behaviour of the action for these classical solutions it
is easy to see that the Euclidean action of the Lorentzian solutions is
pure imaginary.

On the other hand, the  Euclidean solutions have real Euclidean action. 

\item CASE 2 corresponds to $1<\alpha<4/3$.

The situation here is very similar to that of
CASE $1$, the only difference being that now the relevant critical point
where  the Lorentzian classical branch  peaks is $1/4$. As
before we have Lorentzian classical solutions for values of $T$ from $T_{crit}$ all the
way up to $+\infty$, while the existence of  Euclidean classical solutions 
is limited to $T\in[0,T_{crit}(\alpha)]$. However, in this case the
ceiling $T_{crit}(\alpha)$ is much more effective because in the range
$1<\alpha <4/3$, $T_{crit}(\alpha)$ takes low values and there is no
possibility of making it as high as we want by taking the limit
$\alpha\rightarrow 1/3$, as in the previous case. 

In what concerns the action the situation is also similar  to case
$1$. The Euclidean solutions all have real Euclidean action. As for  the
Lorentzian solutions they all have pure imaginary actions.

\item CASE 3 corresponds to $\alpha>1$.

As we can see from figure $6$ the situation here is different. There are two different branches of
classical Lorentzian solutions $\eta_{1}(T,\alpha)$ and $\eta_{2}(T,\alpha)$  peaking respectively at the branch
points $\eta=1/4$ and $\eta=\alpha/4$.  We also  have the usual
classical Euclidean branch,   for $\eta\in[\eta_{0},\infty)$.

This time the Euclidean and Lorentzian regimes are not strictly
separated at the ``boundary'' $T=T_{crit}(\alpha)$. This is because
although the Euclidean solutions are still limited to boundary data
$T<T_{crit}(\alpha)$, there is now one 
branch of Lorentzian solutions $\eta_{2}(T,\alpha)$, that contains
solutions for all positive values  of the boundary
data $T$, including $T<T_{crit}(\alpha)$.

Another big difference is that while the Lorentzian solutions in the 
$\eta_{1}(T,\alpha)$ branch  all have real Euclidean actions,
diverging  to $+\infty$, as $\eta\rightarrow 1/4$, the solutions in
the other Lorentzian branch $\eta_{2}(T,\alpha)$ have fully complex
actions. The most important feature of the Euclidean action of these
solutions is its behaviour near the branch point $\alpha/4$.

$$Im[ I(\eta_{cl}\rightarrow \frac{\alpha}{4}^{-},T,\alpha)]\rightarrow
+\infty ,\,\,\,\,Re[ I(\eta_{cl}\rightarrow \frac{\alpha}{4}^{-},T,\alpha)]\rightarrow
-\infty;$$
the behaviour of $ReI(\eta_{cl})$  can be seen in figure $7$.

As for the Euclidean solutions, their behaviour is similar to that of the
two previous cases. There are real Euclidean solutions with real
Euclidean action for all the
range $[\eta_{0},\infty)$, but they only correspond to positive values
of $T$ for values of $\eta$ slightly larger than $\eta_{0}$, and they
are limited to a maximum value,  $T_{crit}(\alpha)$, of $T$ . In the case of $\alpha=2$, and so $\eta_{0}=0.6$, the
real Euclidean solutions only correspond to $T>0$ from approximately
$\eta=0.631$. 
\end{itemize}

Note that we have Lorentzian classical solutions for the late universe
for all possible values of $\alpha$. To see this remember that  by
late universe we mean very large boundary $3-$spaces, which in our
model corresponds to $a=\alpha (Tl^{2}/H^{2})\rightarrow
+\infty$. However, since we are restricted to $\alpha>1/3$, then
$a\rightarrow +\infty \Leftrightarrow T\rightarrow +\infty$.

\section{Steepest Descents Contour}

In this minisuperspace model the wavefunction of the Universe is

\be
\Psi [  T,\alpha]=\int_{C} D\eta \,e^{-I[\eta, T,\alpha]}   
\label{psi}
\ee

Once more this expression is only heuristic until we choose the
integration contour $C$ and the integration measure $D\eta$. As in
the previous models these choices are largely independent, if we
restrict our attention to polynomial measures like

\be
 D\eta=\frac{ds_{i}}{2\pi il^{2}}=\frac{T}{2\pi iH^{2}}d\eta   
\ee

In accordance with the previous chapters we shall follow Hartle's
prescription for the integration contour, namely, that the correct
integration contour is the steepest descents
contour over complex metrics passing through the dominant extrema of
the Euclidean action. We know that all classical solutions occur in pairs,
$\eta_{cl}(\alpha,T)=\eta_{cl}^{I}(\alpha,T)=\eta_{cl}^{II}(\alpha,T)$,  with Euclidean
actions of opposite sign since they are  located
respectively in the first and second   sheets of the action.
Since by definition all points in a  SD path have the same 
imaginary part of the action we see that no single SD path  can pass
through both extrema. However, if we consider a SD contour made of two
complex conjugate sections one existing on the first sheet the other
on the second ,  since, $\cite{h2}$:

$$ I[\eta]=[I[\eta^{*}]]^{*} $$

and $$ I[\eta_{I}]=- I[\eta_{II}]$$
where $*$ denotes complex conjugation, we see that if one section goes through $\eta_{cl}^{I}(\alpha,T)$ the
other will go through $\eta_{cl}^{II}(\alpha,T)$, and the resulting
wavefunction will be real if the actions are purely imaginary or purely real.

Another way of seeing this is to remember that we are now working for
$\eta\in \bf{R}$, and not $\eta\in {\bkC}$. As such, the SD contour passing
through a value of  $\eta_{cl}(\alpha,T)$ is
 
\be
C_{SD}(T,\alpha)=\biggl\{\eta \in {\bf{R}}  : Im[I(\eta,T,\alpha)]=Im[I(\eta_{cl}(T,\alpha),T,\alpha)]  \biggr\}
\label{csd}
\ee
it is then easy to conclude from our knowledge of the behaviour of
the action in the several sheets of  $\bf{R}$ that if $C_{SD}$ passes through
$\eta_{cl}^{I}$ in ${\bkC} _{1}$ it will pass through
$\eta_{cl}^{II}$in ${\bkC} _{2}$.

For brevity we shall display only the SD contours associated with the classical
solutions in the range $\alpha>4/3$, ie., case $3$, because the contours associated
with the cases $1$ and $2$ are similar to the contours associated with
case $3$.

In figure $8$ we present the SD contour passing through the classical
Lorentzian 
solution $\eta_{cl}^{1}(T=100,\alpha=2)=0.2401$, belonging to the first
branch of Lorentzian solutions that peaks at $\eta=1/4$.

Starting off at the real Lorentzian classical solution
$\eta_{cl}(T=100,\alpha=1)=0.2401$, if we move upwards the SD contour
goes to infinity in the first quadrant along the parabola

\be
\frac{\sqrt{3\alpha-1}}{H^{2}}T[T-T_{crit}^{I}(\alpha)]\sqrt{\eta}=\tilde{I}[\eta_{cl}=0.2401]=265.71
\ee
where $\tilde{I}=Im(I)$.

The convergence of the integral along this part of the SD contour is
guaranteed by the asymptotic behaviour of the action
\be
Re[I^{I}(\eta\rightarrow \infty,\alpha,T)]\sim \frac{\sqrt{3\alpha-1}}{H^{2}}T[T-T_{crit}^{I}(\alpha)]\sqrt{\mid\eta\mid}
\label{as1}
\ee
since for this branch of Lorentzian solutions we always have
$T>T_{crit}^{I}(\alpha)$, and in particular $T^{I}_{crit}(\alpha=2)=17.03$.

If instead we move downwards from the classical solution we
immediately cross onto the second sheet of the action, where the SD
contour encircles the branch point $\eta=1/4$ crossing the branch cut
$(-\infty, \eta_{0}]$ once more, this time between $1/4$ and
$\alpha/4=1/2$, at $\eta=0.2632$. When it does so it emerges onto the third sheet of the
action where it goes to infinity in the first quadrant along the
parabola

\be
\frac{\sqrt{3\alpha-1}}{H^{2}}T[T-T_{crit}^{III}(\alpha)]\sqrt{\eta}=\tilde{I}[\eta_{cl}=0.2401]=265.71
\ee

Once more the convergence of the integral along this part of the SD
contour is guaranteed by the asymptotic behaviour of the action in the
third sheet

\be
Re[I^{III}(\eta\rightarrow \infty,\alpha,T)]\sim \frac{\sqrt{3\alpha-1}}{H^{2}}T[T-T_{crit}^{III}(\alpha)]\sqrt{\mid\eta\mid}
\label{as2}
\ee
since $T_{crit}^{III}$ is always small or negative. In this particular
case  $T^{I}_{crit}(\alpha=2)=-16.69$

The other section of this SD contour passing through the other
extremum located at the second sheet is just the complex conjugate of
this contour and so we will not show it here.

We now study the SD contours associated to the second branch of
classical Lorentzian solutions, i.e., the one that peaks at $\eta=\alpha/4$. 

In figure $9$ we display the SD contour associated with the
Lorentzian classical solution $\eta_{cl}^{2}(T=100,\alpha=0.4931)$. Moving
upwards  we again go to infinity in the first quadrant
of the first sheet along the parabola

\be
\frac{\sqrt{3\alpha-1}}{H^{2}}T[T-T_{crit}^{I}(\alpha)]\sqrt{\eta}=\tilde{I}[\eta_{cl}=0.4931]=165.53
\ee

The convergence of the integral is guaranteed by the asymptotic behaviour
of the action in the first sheet 

\be
Re[I^{I}(\eta\rightarrow \infty,\alpha,T)]\sim \frac{\sqrt{3\alpha-1}}{H^{2}}T[T-T_{crit}^{I}(\alpha)]\sqrt{\mid\eta\mid}
\ee
given that $T_{crit}(\alpha=2)=17.03$.

However, if we move downwards we cross the branch cut and emerge onto
what we have defined as the fourth sheet of the action. There the SD
contour encircles the branch point $\alpha/4=0.5$, and moves upward to
cross the branch cut again between the branch points $\alpha/4=0.5$
and $\eta_{1}(\alpha=2)=0.571$, more specifically at $\eta=0.5175$. By
doing this it moves onto the fifth sheet, where once more it goes to infinity in the first quadrant along the
parabola

\be
\frac{\sqrt{3\alpha-1}}{H^{2}}T[T-T_{crit}^{V}(\alpha)]\sqrt{\eta}=\tilde{I}[\eta_{cl}=0.4931]=165.53
\ee

The convergence of the SD contour along this section is assured by the
asymptotic behaviour of the action in this fifth sheet

\be
Re[I^{V}(\eta\rightarrow \infty,\alpha,T)]\sim \frac{\sqrt{3\alpha-1}}{H^{2}}T[T-T_{crit}^{V}(\alpha)]\sqrt{\mid\eta\mid}
\ee
This is because $T_{crit}^{V}(\alpha)$ is always negative. In this
particular case $T_{crit}^{V}(2)=-46.55$.

In the case of the Euclidean classical solutions they also occur in
pairs but since they both have real Euclidean action they can both lie
in the same SD path because  the imaginary part of the
action is equal for both of them, i.e., zero. 

  In figure $10$, we present  the SD contour
associated with the pair of Euclidean classical solutions, numerically
equal to 

$$\eta_{cl}^{sheetI}(T=13,\alpha=2)=\eta_{cl}^{sheetII}(T=13,\alpha=2)=1.180$$
but located in
the first and second sheets, and so with real Euclidean actions of
opposite sign,
$I[\eta_{cl}^{sheetI}(T=13,\alpha=2)]=-I[\eta_{cl}^{sheetII}(T=13,\alpha=2)]=-4.033$.

We see that starting at the classical solution
$\eta_{cl}^{sheetI}(T=13,\alpha=2)=1.180$, if we move upwards in the
first sheet the SD contour encircles all branch points and crosses the
branch cut  $(-\infty,\eta_{0}=0.6]$ at $\eta=-0.2638$, emerging onto
the second sheet where it  encircles all branch points in the opposite
direction and
arrives at the second classical solution
$\eta_{cl}^{sheetII}(T=13,\alpha=2)=1.180$. If we do another loop we
will cross back onto the first sheet and
will arrive back where we started, i.e., at the classical solution in
the first sheet $\eta_{cl}^{sheetI}(T=13,\alpha=2)=1.180$. So the SD
contour is clearly closed.
Since there are no other critical points of the action in this SD
contour we do not need to worry about the contribution (to the path
integral) of any more
points other than  the two classical solutions.

\section{Semiclassical Wavefunctions}

We have proved that for any classical solution there is always an SD
contour passing through it. Furthermore by analysis of the asymptotic
behaviour of the action we have been able to establish that any path
integral along such SD contours will always be convergent, since none
of them crosses any singularities and the contribution from the
infinities is vanishing.

However this is not enough to prove that the semiclassical
 approximation is always a good approximation of the SD
 wavefunction. In order to do that we must prove that the contributions
 from the regions about the classical solutions are clearly dominant, when
 compared with the contribution from the rest of the SD contour. For
 this to happen the values of $T$, $\alpha$ and $H$ must be such that, locally,
 the integrand, i.e., $\exp{-I}$ is sharply peaked about the
 extrema.

In our case the classical solutions always occur in pairs with actions
of opposite sign. If these
solutions are Lorentzian and have pure imaginary Euclidean actions,
$I[\eta_{cl},T,\alpha]=ImI[[\eta_{cl},T,\alpha]$, 
then for the linear  CPT symmetric wavefunction of the no boundary proposal we
expect the real combination of these two contributions. So the semiclassical wavefunction takes the form

\be
\Psi _{SC}(T,\alpha)\sim \sqrt{\frac{T^{2}}{2\pi H^{4}\mid I^{''}[\eta_{cl}(T,\alpha)]\mid }}2\cos{\biggl\{ImI[\eta_{cl}(T,\alpha),\,\,T,\alpha]\biggr\}}
\ee
where $'$ denotes $d/d\eta$.

This is what happens for semi\-classical approxi\-ma\-tions ba\-sed on the   the Loren\-tzian classical so\-lu\-tions in cases $1$
and $2$, i.e., when $\alpha \in (1/3,1]$, and $\alpha \in [1,4/3]$. It
is also what happens in case $3$,\, $(\alpha>4/3)$, \, for the first
branch of Lorentzian solutions peaking at $\eta=1/4$.
In figures $11$ and $12$ we can see some examples of these wave
functions.

The semiclassical approximation is good when  $\exp{(-I)}$ is sharply
peaked about the extrema, i.e., when
$ImI[\eta_{cl}(T,\alpha)]$ is large. This will always be the  case for the
late universe . If for example we consider $\alpha=1$ then for
the late universe where $T\rightarrow +\infty$
and $\eta_{cl}^{Lor}\rightarrow 1/4$, the action takes the form 
     
\be
ImI_{cl}(T,\alpha=1)\sim \frac{T^{2}}{2H^{2}}=\frac{1}{2}(\Lambda \,b^{2}/l^{2}) 
\ee
So the semiclassical approximation will be very good for large values
of $T$. It will also be the case over the whole range of $T$, (except for $T_{crit}$),
when $H^{2}=\Lambda l^{2}/3$ is sufficiently small as it certainly is
in our late universe.

However we can see in figures  $11$ and $12$ that the  semiclassical
wavefunctions  diverge as $T\rightarrow T_{crit}^{I}$, but
that is only a symptom that the semiclassical approximation breaks
down there, which is signalled by the fact that
$I^{''}[\eta_{cl}(T,\alpha)\rightarrow 0$ when $T\rightarrow T_{crit}^{I}$.

An oscillating wavefunction of the kind in figures $11$ and $12$ predicts classical Lorentzian
spacetime for the late universe, described by the Lorentzian classical
solutions computed above. What does it say about the evolution of the
wormhole? Well since our late universe corresponds to large $3-$boundaries,
which is the case for our classical solutions when $T=bH^{2}/l^{2}
\rightarrow +\infty$ and $a=\alpha b\rightarrow +\infty$, then it
seems that the wormhole throat will grow with the expansion of the
Lorentzian twin universes. So this wavefunction does not predict the
collapse of the wormhole.

When the classical solutions come in pairs of  Euclidean solutions with real
valued actions of opposite sign, as in the regions $\eta\in[\eta_{0},+\infty)$ where 
$0<T<T_{crit}^{I}$, the solution with negative real Euclidean action
will always dominate over its counterpart. In our case 
the action of the  Euclidean solutions in the first sheet is always
negative, 
and so the solutions on the first sheet will dominate over the ones in
the second sheet.

 However, as discussed above, the semiclassical
approximation based on these solutions, will only be valid if the
region near them gives the dominant contribution to the full SD
contour integration. This only happens when  the action is strongly peaked in the SD section around them.
Since we are now limited to $T< T_{crit}^{I}$, this is only true when $H^{2}$ is small.

 And so the asymptotic behaviour of ($\ref{psi}$) for small $H^{2}$ is
 given by the semiclassical approximation associated with the
 Euclidean solution in the first sheet, $\eta_{cl}^{sheetI}(T,\alpha)$

\be
\Psi _{SC}^{Eucl}(T,\alpha)\sim \sqrt{\frac{T^{2}}{2\pi H^{4}\mid I^{''}[\eta_{cl}(T,\alpha)]\mid }}e^{-I[\eta_{cl}^{sheetI}(T,\alpha),\,\,T,\alpha]}
\ee

 The second derivative of
the action vanishes as  $T\rightarrow T_{crit}^{I}$, and so
it will be this term that will dominate the semiclassical
wavefunction near the ``turning point'', $T_{crit}^{I}$, and lead to
its  divergence there. However, when
$H$ is small, the $exp(-I)$ term is dominant everywhere except when
$T$ is really close to $T_{crit}^{I}$. For small $H$ the
real valued Euclidean action   peaks around the Euclidean classical
solution and so the semiclassical approximation is good for almost all
values of $T$, except when we get too close to $T_{crit}^{I}$.

In figure $13$ we see the plot of this semiclassical wavefunction for
$\alpha=2$ and $H=4.9$.
There is a clear peak away from $T_{crit}^{I}$, at $T=13.35$, where the
semiclassical approximation is valid, which seems to indicate a
preferred value of the boundary edges $b=Tl^{2}/H^{2}$ and consequently
 $a=b\alpha=2b$. This peak becomes more pronounced as
$H$ becomes smaller. We thus seem to have a prediction of a favoured
size of the wormhole throat, when we are dealing with wormholes
between Euclidean universes.

\vspace{10pt}
We now discuss a third kind of classical solutions. When
$\alpha>4/3$, we know that there are two branches of Lorentzian
classical solutions, branch 
$1$ peaking at $\eta=1/4$ and branch $2$ peaking at
$\eta=\alpha/4$. These solutions in the second branch have fully
complex Euclidean actions.

$$I[\eta^{cl}_{2}(T,\alpha)]=ReI[\eta^{cl}_{2}(T,\alpha)]+iImI[\eta^{cl}_{2}(T,\alpha)]$$

Their counterparts on the second
sheet will also have complex actions, but of opposite sign. So no
linear combination of the contribution from these two solutions will
ever lead to a real wavefunction. Moreover since the real part of the Euclidean action
is always negative for the classical solutions
$\eta^{cl}_{2}(T,\alpha)$ on the first sheet, even diverging to
$-\infty$, as $\eta\rightarrow \alpha/4$, as can be seen
in figure $7$, it seems obvious that the extremum in the first sheet
will dominate the path integral.  

In order to understand a little better this situation we need to do a
slight diversion, \cite{halliwell}.

\vspace{15pt}

The starting point of a semiclassical/WKB approximation is to assume
that the full wavefunction is well approximated by the contribution of
the regions about some finite number of classical extrema. In our case
it is just one and so the wavefunction can always be written as

\be
\Psi[h_{ij}]=C\exp{\frac{-I[h_{ij}]}{\hbar}}
\ee
where $I=I_{r}+iI_{i}$, where $I_{r}$ and $I_{i}$ are the real and
imaginary parts of the Euclidean action, and the pre-factor $C$
is a slowly varying amplitude containing all higher-order corrections.

At the level of the first approximation in $\hbar$ the Wheeler-DeWitt
equation on $\Psi[h_{ij}]$ is then reduced to the classical
Hamilton-Jacobi equation on $I$, $\cite{banks}$.

\be
H_{0}\biggl(\pi ^{ij}=i\frac{\delta I}{\delta h_{ij}},h_{ij}\biggr)=0
\ee
\vspace{10pt}

\be
\biggl[G_{ijkl}\frac{\delta  I}{\delta h _{ij}}\frac{\delta  I}{\delta h _{kl}}+\sqrt{h}(^{3}R-2\Lambda)\biggr]=0
\ee

\vspace{10pt}

where $G_{ijkl}=\frac{1}{2\sqrt{h}}(h_{ik}h_{jl}+h_{il}h_{jk}-h_{ij}h_{kl})$
and the $\pi ^{ij}$ are the canonical
conjugate momenta of $h_{ij}$.

The action  $I[h_{ij}]$ that solves the the Hamilton-Jacobi  equation is the
classical action of a  congruence of solutions, $[h_{ij}(t)]$, of the classical equations of
motion. These trajectories, parametrised by
their coordinate time $t$, are defined by

\be
\pi^{ij}=i\frac{\delta I}{\delta h_{ij}}
\ee

If the action $I$ is real, the wavefunction has an exponential form and
can be formally interpreted as an ensemble of classical Euclidean
solutions/trajectories. If the action $I$ is pure imaginary, then the wavefunction
has oscillatory behaviour, and describes an ensemble of classical
Lorentzian solutions.

However, if the action $I$ is fully complex the Hamilton-Jacobi
equation can be decomposed into the real part

\be
(\nabla I_{r})^{2}-(\nabla I_{i})^{2}+\sqrt{h}(^{3}R-2\Lambda)=0
\label{jacr}
\ee
and the imaginary part
\be
(\nabla I_{r})\times(\nabla I_{i})=0 
\ee

 From
$\ref{jacr}$, we can see that if the gradient of $I_{i}$ becomes much
larger than the gradient of $I_{r}$, then $I_{i}$ will be an
approximate solution to the Hamilton-Jacobi equation. This means that
the classical evolution is almost Lorentzian.

If $I$ defines an ensemble of almost Lorentzian solutions, then the
real part of the action is also relevant, since it will contribute the
term $\exp(-I_{r})$ to the pre-factor. Since it is exponential it will
probably be the dominant contribution to the pre-factor. Then the
probability measure provided by the pre-factor on the ensemble of
classical solutions is, to leading order, of the form
$\exp(-2I_{r})$, determining the relative weight of different trajectories.

This is exactly what happens in the case of branch $2$ of classical
solutions, $\eta_{2}^{cl}(T,\alpha)$ when $\alpha>4/3$. As can be seen
in figure $14$, the ratio $R$ between the gradients of the imaginary  and
real parts of the Euclidean action

\be
R= \frac{\mid\nabla ImI[\eta=\eta_{2}^{cl}]\mid}{\mid\nabla ReI[\eta=\eta_{2}^{cl}]\mid} 
\ee
along the classical solutions of branch $2$, becomes infinitely large
for the late universe, i.e., for $a=\alpha (Tl^{2}/H^{2})\rightarrow +\infty \Leftrightarrow
\eta_{2}^{cl}\rightarrow {\frac{\alpha}{4}}^{-}$. So for the late
universe $ImI$ really solves the classical Hamilton-Jacobi equation.  
And so the wavefunction should describe an ensemble of Lorentzian
solutions for the late universe.

The contribution of the imaginary part of the action to the
wavefunction does indeed yield the characteristic oscillating
behaviour associated with  the prediction of classical Lorentzian
spacetime for the late universe, as can be seen in figure $15$.

The real part of the action provides the probability measure,  $\exp{\{-2ReI[\eta_{2}^{cl}]\}}$, on the
ensemble of classical Lorentzian universes, determining the relative
weight of these dif\-fer\-ent tra\-jec\-to\-ries. From figure $7$, we see that
 $\exp{\{-2ReI[\eta_{2}^{cl}]\}}$ becomes infinitely large as
$T\rightarrow +\infty$, since $ReI[\eta_{2}^{cl}]\rightarrow -\infty$ as
$\eta_{2}^{cl}\rightarrow {\frac{\alpha }{4}}^{-}$.
It thus seems that in this model,  when $\alpha>4/3$ if we choose the classical solutions
of branch $2$ as the basis for the semiclassical approximation of the
wavefunction of the universe, then configurations consisting of extremely large Lorentzian
classical universes connected by similarly large wormholes 
will dominate.

\section{Conclusions}

The minisuperspace approximation 
 comes very naturally in simplicial quantum gravity,
because by discretising spacetime we are  in effect substituting the
functional $h_{ij}({\bf{x}},t)$, by a set of simple variables, the
edge lengths. The minisuperspace approximation consists only in
restricting all these edge lengths to being equal to a small number of
parameters. So the loss of generality is much smaller than when we
impose the minisuperspace approximation to the continuum. Furthermore,
in the simplicial minisuperspace,  spacetime continues to be treated
in a fully $4-$dimensional way.  There is no need to invoke any
arbitrary $3+1$ decomposition of spacetime, usually ADM, as in the case of
continuum models. This means that global issues like topology can
still be addressed in the simplicial minisuperspace, which is not
possible in the continuum versions. 
The versatility of our approach is evident in the ease in which one
can construct a $4-$D wormhole configuration. Furthermore, since the
signature of the simplicial spacetime  is  only dependent on the ratio
between  edge lengths, we can deal with Lorentzian and
Euclidean configurations at the same time. Indeed with the same model
we were able to study microscopic Euclidean wormholes and large Lorentzian
wormholes.

The simplicial geometry used is quite simple and so it cannot take into account some
relevant dynamics that we would like to model. In particular it would
be better if we could have developed a solvable model with an additional
interior edge length, that would describe the throat of the
wormhole. However, we decided that before we considered that much more
complicated model it would be very useful to determine the results of
the simplest model. And indeed we found several new and interesting
results. For Euclidean wormholes just above the Planck scale, there is
a strongly preferred non-zero throat size. As $H$ becomes smaller the
peak becomes sharper, and we obtain a very stable configuration. Thus the model predicts
that Euclidean wormholes of that scale are stable insofar as that
their throat do not tend to contract leading to the pinching off of the
two universes. 

For large Lorentzian wormhole configurations the situation is
dependent on the value of $\alpha=a/b$. If $\alpha \in [1/3,4/3]$,
then there is only one family of Lorentzian solutions, all with pure
imaginary Euclidean action. The semiclassical wavefunction is a very
good approximation of the full wavefunction for the late, large
Lorentzian universes. Its oscillating behaviour definitely predicts
classical Lorentzian configurations for large universes, with a
similarly large wormhole throat connecting them. However, when
$\alpha>4/3$, a new family of Lorentzian solutions is present. The
fact that their Euclidean action is fully complex, with   a real part
that diverges when the boundary $3-$space becomes large, (i.e. late
universe), seems to indicate that these solutions will dominate the
path integral leading to the full wavefunction. A semiclassical interpretation of the
resulting wavefunction is that it describes an ensemble of classical Lorentzian 
spacetimes, weighted by $\exp{(-2ReI)}$. Since this exponential
diverges to $+\infty$ for increasingly large $3-$spaces, we can
conclude that large  wormhole configurations between very large
Lorentzian universes are very strongly favoured.

\end{document}